\documentclass[runningheads]{svmult}

\usepackage{makeidx}   
\usepackage{graphicx}  
\usepackage{subeqnar}  
\usepackage{multicol}  
\usepackage{physprbb}  
\makeindex             

\usepackage{amsmath}  
\begin{document}
\title*{ Magnetism of correlated systems: beyond LDA
}
\toctitle{ Magnetism of correlated systems: beyond LDA
}
%
%
\titlerunning{Magnetism of correlated systems}
%
\author{A. I. Lichtenstein\inst{1} \and M. I. Katsnelson\inst{2} }
\authorrunning{A. I. Lichtenstein et al.}
%
%
\institute{University of Nijmegen, 6525 ED Nijmegen, The Netherlands
\and Institute of Metal Physics, 620219 Ekaterinburg , Russia} 
\maketitle              

\begin{abstract}
A novel approach to electronic correlations in 
magnetic crystals which takes into account a dynamical many-body
effects is present. In order to to find a 
 frequency dependence of the electron self energy, 
an effective quantum-impurity many-particle problem need to be solved 
within the dynamical mean-field theory.
The numerically exact QMC-scheme and  the spin-polarized fluctuation
exchange approximation are used for the self-consistent solution of this 
single-site many-particle problem.
The calculations of effective exchange interaction
parameters based on the realistic electronic structure of correlated
magnetic crystals have been discussed.
\end{abstract}

\section{Introduction}

The calculation of thermodynamic properties and excitation spectra of
different magnetic materials is one of the most important problems of the
microscopic theory of magnetism.
A modern computational material science is based on the density
functional (DF) approach \cite{lsdf}. It is a common practice to use 
this scheme not only for the  total energy calculations 
and related quantities such as charge and spin densities, 
but also for different spectral characteristics. Sometimes an agreement
of the computational results with the experimental data is very impressive,
despite the absence of a reliable theoretical background. In principle, 
the energies of Kohn-Sham quasiparticlies \cite{lsdf} which are calculated
in a standard  band theory are just auxiliary quantities for 
the total energy calculation. 
Among all achievements of the quantitative electronic theory
a list of difficulties and shortcomings grows, especially when one consider
the  magnetic d- and f- electron systems. In a number of cases the theory 
appeared to be {\it qualitatively} inadequate. First, the DF scheme cannot describe
correctly the phenomenon
of ``Mott insulators'' \cite{mott}, as it was first shown by Terakura {\it {et al}} 
\cite{TWOK}
in attempt to calculate the
electronic structure of 3d-metal oxides. Later we faced  similar
problems in field of the high-T$_c$ superconductors \cite{pickett} and
other compounds \cite{revU}. The Ce- and U-
based ``heavy fermion'' compounds such as CeCu$_6$, UPt$_3$, etc, are another
``hot-spots'':  normally the calculated values of effective masses are  
orders of magnitude smaller than the experimentally observed ones \cite{zwicknagl}.
Even for the pure 3d metals
some qualitative differences between the theory and experiments exists.
For example, there are at least three difficulties with the  
photoelectron spectra of ferromagnetic nickel \cite{eastman}:
(i) the measured width of the occupied
part of d-band is 30$\%$ narrower than the calculated one (ii) the
spin splittings is twice smaller than in the LSDA 
and (iii) the band structure cannot describe a famous 6 eV - satellite. Calculations
for paramagnetic spin-disordered states \cite{gyorffy}
lead to the conclusion that Ni has no local magnetic moments above the
Curie temperature T$_c$, in a clear contradiction with experimental results 
\cite{niloc}. For iron, a standard band
theory cannot explain the data about spin polarization of thermionic
electrons \cite{thermoem,monnier,lda+sp} and some features of angle-resolved
photoemision spectra \cite{satyes,kirby,lda+sp}. 
All these difficulties put many questions to 
the DF approach: what are the ``electron spectrum'' which we really
calculated and how one can improve the electronic theory for magnetic 
d- and f- systems?

It was understood many years ago that all this problems are connected
with inadequate description of many-body effects in DF calculations
of the excitation spectra. Such methods as GW \cite{gw},  LDA+U \cite{revU}  and SIC
(Self-Interaction Correction approximation) \cite{sic} have been proposed
to improve the situation. This methods are very useful for the
description of antiferromagnetic transition-metal oxides as the Mott insulators \cite{sic,revU}. However,
one should note that both LDA+U and SIC are just
the mean-field approximations
and cannot describe the {\it
correlation} effects which are, by definition, the many-body effects  {\it
beyond} the Hartree-Fock. For example, in these approaches one
need spin or orbital ordering to describe the Mott insulator 
and it is impossible to describe correctly electronic structure
of NiO or MnO in paramagnetic phase. At the same time, the
magnetic ordering should not be important for the basic physics of
Mott insulators \cite{mott}. All the ``Hartree-Fock-like'' approaches
fail to describe the renormalization of the effective mass
in the heavy fermion systems. There are also many problems concerning
the electronic structure and itinerant magnetism of 3d metals as described above.
Thus, one need some practical ways to incorporate {\it correlation} effects in the
electronic structure of solids.

In principle, there are two ways to include them into DF calculations.
The first one is the use of {\it time-dependent} DF formalism which
can guarantee, in principle, an opportunity to calculate exact response
functions \cite{timedep}, in the same sense as the Hohenberg-Kohn theorem
guarantees the total energy in usual ``static''
DF \cite{lsdf}. However, all the expressions for this time-dependent
non-local DF in real calculations are based on
RPA-like approximations which described not satisfactory the really
highly correlated systems. They are excellent for investigation the
plasmon spectrum of aluminum, but not for understanding a 
nature of high-$T_{c}$ superconductivity or the heavy fermion behavior.
Another way is to use an ``alternative'' many-body theory developed
in the 50-th by Gell-Mann and Brueckner, Galitskii and Migdal,
Beliaev and many others in terms of the Green functions rather than in
the electron density \cite{AGD}. We try to formulate such computational 
approach as a generalization of LDA+U scheme, a so-called ``LDA++'' method.
The main difference of LDA++ approach from LDA+U one is an account
in the former dynamical fluctuations, or the real correlation effects,
described by local but energy dependent self-energy $\Sigma(E)$,
so the LDA++ means $LDA+U+\Sigma(E)$.

The comparison of the standard DF theory in the 
local spin density approximation (LSDA) and LDA++ approach
is represented in the table I.

\begin{table}
\caption{Comparison of different schemes}
\begin{center}
\renewcommand{\arraystretch}{1.4}
\setlength\tabcolsep{5pt}
\begin{tabular}{ll}
\hline
\noalign{\smallskip}
LSDA & LDA++ \\ 
\noalign{\smallskip}
\hline
\noalign{\smallskip}
Density functional & Baym-Kadanoff functional \\ 
Density  $\rho (\mathbf{r}) $ & Green-Function $G(\mathbf{r,r}^{\prime },E)$ \\ 
Potential  $ V_{xc}(\mathbf{r})$ & Self-energy $\Sigma _{i}(E)$ \\ 
$E_{tot}=E_{sp}-E_{dc}$ & $\Omega =\Omega _{sp}-\Omega _{dc}$ \\ 
$E_{sp}=\sum_{\lambda <\lambda _{F}}\varepsilon _{\lambda }$ & $\Omega
_{sp}=-Tr\ln [-G^{-1}] $\\ 
$E_{dc}=E_{H}+\int \rho V_{xc}d{\mathbf{r}}-E_{xc}$ & $\Omega _{dc}=Tr\Sigma
G-\Phi _{LW}$ \\ 
\noalign{\smallskip}
\hline
\end{tabular}
\end{center}
\label{Tab1a}
\end{table}
First of all the LSDA theory is based on the Hohenberg - Kohn theorem
that the total energy $E_{tot}$ is a functional of charge and
spin densities, while the
LDA++ scheme considers the thermodynamic potential $\Omega$ as a functional of exact one-particle
Green
functions. This approach in many-particle theory has been introduced  in
the works by
Luttinger and Ward \cite{luttinger} and Baym and Kadanoff \cite{BK}. The Green
function
in LDA++ theory plays the same role as the density matrix in LSDF
formalism.
We stress the dynamic nature of the correlation effects which are taken
into account in
the LDA++ approach since the density in the LSDA is just the static limit of
the local Green function.
Further, the self energy $\Sigma$ is
analogous to the exchange-correlation potential; local approximation
for  $\Sigma$,
which is assumed to be energy-dependent but not momentum-dependent
corresponds
to the local approximation for $V_{xc}$. In both formalisms the
thermodynamic potential
can be represented as a ``single-particle'' one, $\Omega_{sp}$ minus the
contributions
of the ``double counted terms'', $\Omega_{dc}$. It will be important for
the consideration
of so-called ``local force theorem'' and the computation of magnetic
interaction parameters
(Sect. III). The single-particle contribution to the
thermodynamic potential
in the LDA++ would have the same form as in the LSDA if we take into account only
a pole part of
the Green function and neglect the quasiparticle damping. However, even
in that case
the quasiparticle energies are not the same since the poles of the Green
functions are
not coincide, generally speaking, with the ``Kohn-Sham'' energies. The
quantity $\Phi_{LW}$
is the Luttinger-Ward generating functional for
the self energy, or
the sum of all the skeleton diagrams without free legs \cite{luttinger}.

The difficulty with the  finite temperature effects is one
of the main
shortcomings of a standard DF formalism. In all realistic calculations
the temperature
is included in the Fermi distribution functions and in the
lattice constants
via the thermal expansion \cite{TLDA}
. At the same time,
for the itinerant
electron magnets the temperature effects connected with the ``Bose''
degrees of
freedom due to spin waves and paramagnons are much more important \cite{moriya}.
In principle, these effects could be taken into account in the DF theory via
the temperature
dependence of the exchange-correlation potential, the corresponding
terms being
 nonlocal. It is not easy to propose an adequate
expression for
such temperature-dependent non-local potential. 
One of the first attempts
in this direction based on simple RPA-like considerations \cite{kotani}.
On the other hand, in LDA++ type of scheme all the calculations are naturally carried
out for finite
temperatures by the using of Matsubara frequencies, as in the usual 
many-body
theory \cite{AGD}.

The main assumption of the LDA++ approach is the importance of only
intrasite
``Hubbard correlations'' with the local approximation for the self-energy.
It is
worthwhile to stress a difference of this kind of locality from the
locality in
DF theory. In the latter, the local approximation means that the
exchange-correlation
energy is calculated for the homogeneous electron gas \cite{ceperley}.
It is known from exact QMC calculations that the correlation
effects
could lead to some instabilities of the state of homogeneous electron
gas (magnetism,
charge ordering, etc) only for electron densities which are  order of 
magnitude smaller than ones typical for real metals (the critical
values of
the parameter $r_s$ are of order of hundred in comparison with the
``normal'' range 2-6 for metals).
At the same time, magnetism and charge ordering are rather usual for
real
compounds with the d- and f- elements. It seems that the ``atomic-like''
features
of d- and f- states are of the crucial importance to describe the
correlation effects
in real compounds. Only these features are taken into account in the
Hubbard-like terms for the d- or f-states 
in LDA++ approach. Therefore one can view the LDA++ as a simplest
way for quantitative considerations of the correlation effects in
the transition metals,
and their compounds, based on the LSDA description for all non-correlated
electrons in the systems.

The investigation of correlation effects in the electronic structure and magnetism
of iron-group metals is still far from the final picture and attracts continuous
interest (see, e.g., \cite{gyorffy,liebsch,steiner,IKT} and Refs therein). 
Despite of 
many attempts, the situation is still unclear both theoretically and
experimentally. For example, there is no agreement on the presence of 5 eV
satellite in photoemission spectrum of iron \cite{satyes,kirby}, and on the
existence of local spin splitting above Curie temperature in nickel \cite
{splitni}. 
From the theoretical point of view, different
approaches such as the second-order perturbation theory \cite{treglia,steiner},
the T-matrix approximation \cite{liebsch,drhal},
the three-body Faddeev approximation \cite{manghi},
and the  moment expansion method \cite{nolting} were used.
Unfortunately, the  applicability of these
schemes are not clear. Here we present the version of ``LDA++'' 
approach \cite{lda+sp,lda++,exchlda} which is based on the combination
of standard band theory technique with so-called dynamical mean-field
theory (DMFT) or LDA+DMFT scheme \cite{poter}. 

\section{Computational technique}

We start from LDA+U Hamiltonian in the diagonal density approximation: 
\begin{eqnarray}
H &=&\sum_{\{im\sigma \}}t_{im,i^{\prime }m^{\prime }}^{LDA}c_{im\sigma
}^{+}c_{i^{\prime }m^{\prime }\sigma }+  
\frac 12\sum_{imm^{\prime }\sigma }U_{mm^{\prime }}^in_{im\sigma
}n_{im^{\prime }-\sigma }+  \nonumber \\
&&\frac 12\sum_{im\neq m^{\prime }\sigma }(U_{mm^{\prime }}^i-J_{mm^{\prime
}}^i)n_{im\sigma }n_{im^{\prime }\sigma } \label{calc1} 
\end{eqnarray}
where $i$ is the site index and $m$ is the orbital
quantum numbers; $\sigma =\uparrow ,\downarrow $ is the spin projection; $%
c^{+},c$ are the Fermi creation and annihilation operators ($n=c^{+}c$);
$t^{LDA}$ is effective single-particle Hamiltonian obtained from the
non-magnetic LDA with the correction for double counting of the average
interactions among d-electrons. In the general case of spin-polarized LSDA
Hamiltonian this correction is presented in Refs. \cite{lda++,lda+sp,lda+u}.
In the non-magnetic LDA this is just a shift ``back'' of correlated d-states
with respect to s,p-states by the average Coulomb and exchange potential:
$\Delta_d=U(n_d-\frac 12)-\frac 12J(n_d-1)$, where U and J are the average values of
$U_{mm^{\prime }}$ and $J_{mm^{\prime}}$ matrices and $n_{d}$ is the average number of
d-electrons.

The screened Coulomb and exchange vertex for the d-electrons{\ 
\begin{eqnarray}
U_{mm^{\prime }} =<mm^{\prime }|V_{scr}^{ee}({\bf r-r}^{\prime
})|mm^{\prime }> \smallskip 
, \smallskip 
J_{mm^{\prime }} =<mm^{\prime }|V_{scr}^{ee}({\bf r-r}^{\prime
})|m^{\prime }m> \label{hamu} 
\end{eqnarray}
are expressed via the effective Slater integrals.
 We use the minimal $spd$%
-basis in the LMTO-TB formalism \cite{OKA} and numerical orthogonalization for
 $t^{LDA}({\bf k})$  matrix \cite{lda++}. 
Local density approximation \cite{lsdf} was used for 
the self-consistent electronic structure calculation.

In order to find the best local approximation for the self-energy we use the DMFT
method \cite{dinf} for real systems. This scheme become exact in the limit of 
infinite lattice coordination number \cite{metzner}.
The DMFT approach reduce the lattice many-body problem (Eq.(\ref{hamu})) to the
self-consistent solution of effective one-cite Andersen model.
In this case we need a 
local Green-function matrix which has the following form in the orthogonal LMTO-representation: 
\begin{equation}
G(i\omega )=\sum_{{\bf k}}\{i\omega +\mu -t^{LDA}({\bf k})-\Sigma (i\omega
)\}^{-1}  \label{Gk}
\end{equation}
were $\mu$ is the chemical potential. 
Note that due to cubic crystal symmetry of ferromagnetic bcc-iron the local
Green function without spin-orbital interactions is diagonal
both in the orbital and the spin indices. The so-called
bath Green function which defined the effective Andersen model
and  preserve the double-counting of the local
self-energy is obtained as a solution of the following impurity model \cite{dinf}: 
\begin{equation}
G_{0m}^{-1}(i\omega )=G_m^{-1}(i\omega )+\Sigma _m(i\omega )  \label{Sig}
\end{equation}

The local Green functions for the imaginary time interval $\left[ 0,\beta
\right] $ with the mesh $\tau _l=l\Delta \tau $, $l=0,...,L-1,$ and $\Delta
\tau =\beta /L$, where $\beta =\frac 1T$ is calculated in the path-integral
formalism \cite{dinf,MQMC}:{\ 
\begin{equation}
G_m^{ll^{\prime }}=\frac 1Z\sum_{s_{mm^{\prime }}^l}\det
[O(s)]*G_m^{ll^{\prime }}(s)  \label{QMC}
\end{equation}
here we redefined for simplicity $m\equiv \{m,\sigma \},Z$ is the partition
function and the so-called fermion-determinant }$\det [O(s)]$ 
as well as the Green
function for arbitrary set of the auxiliary fields{\ }$G(s)=O^{-1}(s)$ {are
obtained via the Dyson equation \cite{hirsch} for imaginary-time matrix} $(%
{\bf G}_m(s)\equiv G_m^{ll^{\prime }}(s))$: 
\[
{\bf G}_m=[{\bf 1}-({\bf G}_m^0-{\bf 1})(e^{V_m}-{\bf 1})]^{-1}{\bf G}_m^0 
\]
{where the effective fluctuation potential from the Ising fields }$%
s_{mm^{\prime }}^l=\pm 1$ is

{\ 
\begin{eqnarray*}
V_m^l  = \sum_{m^{\prime }(\neq m)}\lambda _{mm^{\prime }}s_{mm^{\prime
}}^l\sigma _{mm^{\prime }} \, , \, {\textrm {where}}  \, \,
\sigma _{mm^{\prime }} = \{ 
\begin{array}{c}
1,m<m^{\prime } \\ 
-1,m>m^{\prime }
\end{array}
\end{eqnarray*}
and the discrete Hubbard-Stratonovich parameters are $\lambda _{mm^{\prime
}}={\rm arccosh}[\exp (\frac 12\Delta \tau U_{mm^{\prime }})] $ \cite{hirsch}.
Using the output local Green function from QMC and input bath Green functions
the new self-energy is obtain via Eq.(\ref{Sig}) and the self-consistent
loop can be closed through Eq.(\ref{Gk}). 
The main problem of the multiband QMC formalism is the large number of the
auxiliary fields }$s_{mm^{\prime }}^l.$ For each time slices $l $ it is
equals to $M(2M-1)$ where $M$ is the total number of the orbitals which gives
45 Ising fields for the d-states case.
We compute the sum over this auxiliary fields in
Eq.(\ref{QMC}) using important sampling QMC algorithm and performed a dozen of
self-consistent iterations over the self-energy Eqs.(\ref{Gk},\ref{Sig},\ref
{QMC}). The number of QMC sweeps was of the order of 10$^5$ on the CRAY-T3e
supercomputer.
The final $G_m(\tau )$ has very little statistical noise. We use maximum
entropy method \cite{MEM} for analytical continuations of the QMC Green
functions to the real axis. Comparison of the total density of states (DOS) with
the results of LSDA calculations (Fig.\ref{DOS}) shows a reasonable
agreement for single-particle properties of not ``highly correlated''
ferromagnetic iron.
 We calculate the bcc iron at experimental
lattice constant with 256 {\bf k}-points in the irreducible part of Brillouin zone.
The Matsubara frequencies summation corresponds to the temperature of about T=850 K.
The average magnetic moment is about 1.9 $\mu_B$ which corresponds to a small
reduction of the LSDA-value of 2.2 $\mu_B$  for such a high temperature. 
The DOS curves in the LDA+$\Sigma$ approach with
exact QMC solution of on-site multiorbital problem is similar to that
obtained within the simple perturbative fluctuation-exchange (FLEX)
approximation described below. The discussion of the results 
and the comparison with the experimental data will be given in Section 4.

\begin{figure}
\includegraphics[width=.4\textwidth]{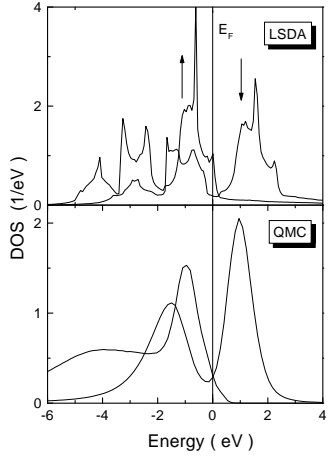}
\vskip  0.5cm
\caption{
Spin-resolved density of d-states and magnetic moments for ferromagnetic 
iron in the LSDA and the LDA+QMC calculations
for different average Coulomb interactions with J=0.9 eV
and temperature T=1500 K.  
}
\label{DOS}
\end{figure}

The QMC method described above is probably the most rigorous real way
to solve an effective impurity problem in the framework of DMFT theory.
However, it is rather time consuming. Besides that, in the previous
section we did not work with {\it complete} four-indices Coulomb
matrix:

\begin{equation}
\left\langle 12\left| v\right| 34\right\rangle =\int d{\bf r}d{\bf r}%
^{\prime }\psi _1^{*}({\bf r})\psi _2^{*}({\bf r}^{\prime })v_{scr}\left( 
{\bf{r-r}}^{\prime }\right) \psi _3({\bf r})\psi _4({\bf r}^{\prime
}),  \label{coulomb}
\end{equation}
where we define for simplicity $m_1\equiv 1$. 

 For moderately strong correlations (which is the case of
iron group metal) one can propose an approximate scheme which is
more suitable for the calculations. It is based on the fluctuation
exchange (FLEX) approximation by Bickers and Scalapino
\cite{FLEX} generalized to multiband spin-polarized case \cite{lda++,lda+sp,mbick}. 
The electronic self-energy in the FLEX is equal to: 

\begin{equation}
\Sigma =\Sigma ^{HF}+\Sigma ^{(2)}+\Sigma ^{(ph)}+\Sigma ^{(pp)} ,
\label{sig}
\end{equation}
where the Hartree-Fock contribution has a standard form:

\begin{equation}
\Sigma _{12,\sigma }^{HF}=\sum_{34}\left[ \left\langle 13\left| v\right|
24\right\rangle \sum_{\sigma ^{\prime }}n_{34}^{\sigma ^{\prime
}}-\left\langle 13\left| v\right| 42\right\rangle n_{34}^\sigma \right] ,
\label{shf}
\end{equation}
with the occupation matrix $n_{12}^\sigma=G_{21}^\sigma (\tau \rightarrow -0)$;
this contribution to $\Sigma$ is equivalent to spin-polarized
``rotationally-invariant'' LDA+U method \cite{lda+u}.

The second-order contribution in the spin-polarized case reads:
\begin{eqnarray}
\Sigma _{12,\sigma }^{(2)}(\tau ) =-\sum_{\left\{ 3-8\right\}
}\left\langle 13\left| v\right| 74\right\rangle G_{78}^\sigma (\tau)* \nonumber \\
*\left[ \left\langle 85\left| v\right| 26\right\rangle \sum_{\sigma ^{\prime
}}G_{63}^{\sigma ^{\prime }}(\tau )G_{45}^{\sigma ^{\prime }}(-\tau )  
-\left\langle 85\left| v\right| 62\right\rangle G_{63}^\sigma (\tau
)G_{45}^\sigma (-\tau ) \right] , 
\label{sig2} 
\end{eqnarray}
and  the higher-order particle-hole (or particle-particle) contribution
\begin{equation}
\Sigma _{12,\sigma }^{(ph)}\left( \tau \right) =\sum\limits_{34,\sigma
^{\prime }}W_{13,42}^{\sigma \sigma ^{\prime }}\left( \tau \right)
G_{34}^{\sigma ^{\prime }}\left( \tau \right)   \label{sigph},
\end{equation}
with p-h (p-p) fluctuation potential matrix:

\begin{equation}
W^{\sigma \sigma ^{\prime }}\left(i\omega \right) =\left[ 
\begin{array}{cc}
W^{\uparrow \uparrow }\left(i\omega \right) & W^{\uparrow \downarrow }\left(i
\omega \right) \\ 
W^{\downarrow \uparrow }\left(i\omega \right) & W^{\downarrow \downarrow
}\left(i\omega \right)
\end{array}
\right] ,  \label{wpp}
\end{equation}
where the spin-dependent effective potentials has a generalized RPA-form and
can be found in \cite{lda+sp}. 
Note that for both p-h and p-p channels the effective interactions,
according to Eq.(\ref{wpp}), are non-diagonal matrices in {\it spin
space}
as in the QMC-scheme, in contrast with any mean-field approximation like LSDA.
This can be important for the spin-dependent transport phenomena
in transition metal multilayers.

We could further reduce the computational procedure by neglecting  dynamical
interaction in the p-p channel since the most important fluctuations in
itinerant electron magnets are spin-fluctuations in the p-h channel. We take into
account static ( of $T$- matrix type) renormalization of effective
interactions  replacing  the bare matrix $U_{12,34}=$ $%
\left\langle 12\left| v\right| 34\right\rangle $ in FLEX-equations 
with the corresponding spin-dependent scattering $T-$matrix

\begin{equation}
\left\langle 12\left| T^{\sigma \sigma ^{\prime }}\right| 34\right\rangle
=\left\langle 12\left| v\right| 34\right\rangle
-\sum\limits_{5678}\left\langle 12\left| v\right| 56\right\rangle
\int\limits_0^\beta d\tau G_{56}^\sigma \left( \tau \right) G_{78}^{\sigma
^{\prime }}\left( \tau \right) \left\langle 78\left| T^{\sigma \sigma
^{\prime }}\right| 34\right\rangle  \label{tmatrix}
\end{equation}

Similar approximation has been checked for the Hubbard model \cite{fleck}
and appeared to be accurate enough for not too large $U$.
Finally, in the spirit of DMFT-approach $\Sigma = \Sigma [G_0]$, and
all the Green functions in the self-consistent FLEX-equations are in fact
the bath Green-functions $G_0$. 
 
\section{Exchange interactions}

An useful scheme for analyses of exchange interactions in the LSDF
approach is a so called ''local force theorem''. In this case  the
calculation of small total energy change reduces to variations
of the one-particle
density of states \cite{MA,LKG}. First of all, let us prove the analog of
local force theorem in the LDA++ approach. In contrast with the standard
density functional theory, it deals with the real dynamical quasiparticles
defined via Green functions for the correlated electrons rather than with
Kohn-Sham ``quasiparticles'' which are, strictly speaking, only auxiliary
states for the total energy calculations. Therefore, instead of the working with
the thermodynamic potential as a {\it density} functional we have
to start from the general expression for  $\Omega $ in terms of 
exact Green function in the Table I. 
We have to keep in mind also Dyson equation 
$
G^{-1}=G_0^{-1}-\Sigma  
$
and the variational identity
$
\delta \Phi_{LW} =Tr\Sigma \delta G.  
$
Here $Tr=Tr_{\omega iL\sigma }$ is the sum over Matsubara frequencies $Tr_\omega
...=T\sum\limits_\omega ...,$ $\omega =\pi T\left( 2n+1\right) ,$ $n=0,\pm
1,...,$ $T$ is the temperature, and $iL\sigma $ are site numbers ($i$),
orbital quantum numbers ($L={l,m}$) and spin projections $\sigma $ ,
correspondingly. 
We represent the expression for $\Omega$ as a difference of ''single
particle'' ($sp$) and ''double counted'' ($dc$) terms as it is usual in the
density functional theory. When neglecting the quasiparticle damping, $%
\Omega _{sp}$ will be nothing but the thermodynamic potential of ''free''
fermions but with exact quasiparticle energies. Suppose we change the
external potential, for example, by small spin rotations. Then the variation
of the thermodynamic potential can be written as 
\begin{equation}
\delta \Omega =\delta ^{*}\Omega _{sp}+\delta _1\Omega _{sp}-\delta \Omega
_{dc}  \label{var2}
\end{equation}
where $\delta ^{*}$ is the variation without taking into account the change
of the ''self-consistent potential'' (i.e. self energy) and $\delta _1$ is
the variation due to this change of $\Sigma $. 
 To avoid  a possible
misunderstanding, note that we consider the variation of $\Omega $ in the
general ``non-equilibrium'' case when the torques acting on spins are nonzero
and therefore $\delta \Omega \neq 0$. In order to study the response of the system
to general spin rotations one can consider either variations of the spin directions at
the fixed effective fields or, vice versa, rotations of the effective fields,
i.e. variations of $\Sigma$, at the fixed magnetic moments. We use the second
way. 
Taking into account the variational property of  $\Phi$ 
one can be easily shown (cf. Ref. \cite{luttinger}) that 
$
\delta _1\Omega _{sp}=\delta \Omega _{dc}=TrG\delta \Sigma  \label{var3}
$
and hence 
\begin{equation}
\delta \Omega =\delta ^{*}\Omega _{sp}=-\delta ^{*}Tr\ln \left[ \Sigma
-G_0^{-1}\right]  \label{var4}
\end{equation}
which is an analog of the ''local force theorem'' in the density functional
theory \cite{LKG}.

Further considerations are similar to the corresponding ones in LSDF
approach. 
In the LDA++ scheme, the self energy is local, i.e. is diagonal in site
indices. Let us write the spin-matrix structure of the self energy and Green
function in the following form 
\begin{eqnarray}
\Sigma _i =\Sigma _i^c+{\pmb \Sigma}_i^s{\pmb { \sigma }}\, , \, \, 
G_{ij} =G_{ij}^c+{\bf G}_{ij}^s{\pmb { \sigma }} \label{spin}  
\end{eqnarray}
where $\Sigma _i^{\left( c,s\right) }=\frac 12\left( \Sigma _i^{\uparrow
}\pm \Sigma _i^{\downarrow }\right)$, ${\bf \Sigma}_i^s=\Sigma _i^s{\bf e}_i,
$ with ${\bf e}_i$ being the unit vector in the direction of effective
spin-dependent potential on site $i$, ${\pmb { \sigma }}=(\sigma_x,\sigma_y,%
\sigma_z)$ are Pauli matrices, $G_{ij}^c=\frac 12Tr_\sigma (G_{ij})$ and $%
{\bf G}_{ij}^s=\frac 12Tr_\sigma (G_{ij} {\bf {\sigma}})$. We assume that
the bare Green function $G^0$ does not depend on spin directions and all the
spin-dependent terms including the Hartree-Fock terms are incorporated in
the self energy. Spin excitations with low energies are connected with the
rotations of vectors ${\bf e}_i$:
$
\delta {\bf e}_i=\delta {\pmb \varphi}_i\times {\bf e}_i  \label{rot1}
$
According to the ''local force theorem'' (\ref{var4}) the corresponding
variation of the thermodynamic potential can be written as 
$
\delta \Omega =\delta ^{*}\Omega _{sp}={\bf V}_i\delta {\bf \varphi}_i
$
where the torque is equal to 
\begin{equation}
{\bf V}_i=2Tr_{\omega L}\left[ {\pmb \Sigma }_i^s\times {\bf G}_{ii}^s\right]
\label{torque}
\end{equation}

Using the spinor structure of the Dyson equation one can write the Green function
in this expression in terms of pair contributions (a similar trick has been proposed in
Ref. \cite{physica} in the framework of LSDF approach). As a result,
we represent the total thermodynamic potential of spin rotations or the
effective Hamiltonian in the form \cite{exchlda} 
\begin{equation}
\Omega _{spin}=-\sum_{ij}Tr_{\omega L}\left\{ \left( {\bf G}_{ij}^s{\bf %
\Sigma }_j^s\right) \left( {\bf G}_{ji}^s{\pmb \Sigma }_i^s\right) -{\pmb %
\Sigma }_i^sG_{ij}^c{\pmb \Sigma }_j^sG_{ji}^c-i\left( {\pmb \Sigma }%
_i^s\times G_{ij}^c{\pmb \Sigma }_j^s\right) {\bf G}_{ji}^s\right\}
\label{Hamilt}
\end{equation}
one can show by direct calculations that 
$
\left[ \frac{\delta \Omega _{spin}}{\delta {\pmb \varphi }_i}\right]
_{G=const}={\bf V}_i  \label{torque3}
$
This means that $\Omega _{spin}\left\{ {\bf e}_i\right\} $ is the effective
spin Hamiltonian. The last term in Eq.(\ref{Hamilt}) is nothing but
Dzialoshinskii- Moriya interaction term. It is non-zero only in relativistic
case where ${\pmb \Sigma }_j^s$ and ${\bf G}_{ji}^s$ can be, generally
speaking, ``non-parallel'' and $G_{ij}\neq G_{ji}$ for the crystals without
inversion center. 

In the nonrelativistic case one can rewrite the spin Hamiltonian for small
spin deviations near collinear magnetic structures in the following form 
$
\Omega _{spin}=-\sum_{ij}J_{ij}{\bf e}_i{\bf e}_j  \label{heis}
$
where 
\begin{equation}
J_{ij}=-Tr_{\omega L}\left( \Sigma _i^sG_{ij}^{\uparrow }\Sigma
_j^sG_{ji}^{\downarrow }\right)  \label{Jij}
\end{equation}
are the effective exchange parameters. This formula generalize the LSDA
expressions of \cite{LKG} to the case of correlated systems.

Spin wave spectrum in ferromagnets can be considered both directly from the
exchange parameters or by the consideration of the energy of corresponding
spiral structure (cf. Ref. \cite{LKG}). In nonrelativistic case when the
anisotropy is absent one has 
\begin{equation}
\omega _{{\bf q}}=\frac 4M\sum\limits_jJ_{0j}\left( 1-\cos {\bf qR}_j\right)
\equiv \frac 4M[J(0)-J({\bf q})]  \label{om}
\end{equation}
where $M$ is the magnetic moment (in Bohr magnetons) per magnetic ion.

It should be noted that the expression for spin stiffness tensor $D_{\alpha
\beta }$ defined by the relation 
$
\omega _{{\bf q}}=D_{\alpha \beta }q_\alpha q_\beta   \label{D}
$
(${\bf q\rightarrow 0}$) in terms of exchange parameters has to be exact
as the consequence of phenomenological Landau- Lifshitz equations
which are definitely correct in the long-wavelength limit. Direct
calculation basing on variation of the total energy under spiral spin
rotations (cf. Ref. \cite{LKG}) leads to the following expression  
\begin{equation}
D_{\alpha \beta }=-\frac 2MTr_{\omega L}\sum\limits_{{\bf k}}\left( \Sigma ^s%
\frac{\partial G^{\uparrow }\left( {\bf k}\right) }{\partial k_\alpha }%
\Sigma ^s\frac{\partial G^{\downarrow }\left( {\bf k}\right) }{\partial
k_\beta }
\right)   \label{D21}
\end{equation}
were ${\bf k}$ is the quasimomentum and the summation is over the Brillouin
zone. 
The expressions Eqs.(\ref{Jij}) and (\ref{om}) 
are reminiscent of usual RKKY indirect exchange interactions in the s-d
exchange model (with $\Sigma^s$ instead  of the s-d exchange integral).

We prove in the Appendix that the expression for the stiffness is exact within the local
approximation.  At the same time, the exchange parameters themselves,  
generally speaking, differ
from the exact response characteristics defined via static susceptibility
since the latter contains vertex corrections. 
The derivation of approximate exchange
parameters from the variations of thermodynamic potential
can be useful for the estimation of $J_{ij}$  
in the different magnetic systems.

\section{Computational results }

We have started from the spin-polarized LSDA band structure of ferromagnetic
iron within the TB-LMTO method \cite{OKA} in the minimal $s,p,d$ basis set and
used numerical orthogonalization to find the $H_t$ part of our starting
Hamiltonian. We take into account Coulomb interactions only between $d$-states.
 The correct parameterization of the $H_{U}$ part is indeed a serious problem.
For example, first-principle estimations of average Coulomb
interactions (U) \cite{U,steiner} in iron lead to unreasonably large value of order of
5-6 eV in comparison with experimental values of the U-parameter
in the range of 1-2 eV \cite{steiner}. 
Semiempirical analysis of the appropriate  interaction value 
\cite{oles} gives $U\simeq 2.3$ eV. The difficulties with choosing  the
correct value of $U$ are connected with complicated screening problems,
definitions of orthogonal orbitals in the crystal, and contributions of the
intersite interactions. In the quasiatomic (spherical) approximation the
full $U$-matrix for the $d-$shell is determined by the three parameters
$U, J$ and $\delta J$ or equivalently by effective Slater
integrals F$^0$, F$^2$ and F$^4$ \cite{lda++,revU}. For example,
U= F$^0$,  J=(F$^2$+F$^4$)/14 and we
use the simplest way of estimating  $\delta J$ or  F$^4$ keeping the ratio 
 F$^2$/F$^4$  equal to its atomic value 0.625 \cite{solj}. 

Note that the value of intra-atomic (Hund) exchange
interaction $J$ is not sensitive to the screening  and  approximately
equals to  0.9 eV in different estimations \cite{U}. 
 For  the most
important parameter $U$, which defines the bare vertex
matrix Eq.(\ref{coulomb}),
we use the value $U=2.3$ eV for Fe \cite{oles}, $U=3$ eV for Co and Mn and $U=4$ eV 
for Ni and Cu.
To calculate the spectral functions:
$
A_\sigma \left( {\mathbf k},E\right) =-\frac 1\pi Tr_LG_\sigma \left({\mathbf%
k},E+i0\right) 
$
and DOS as their sum over the  Brillouin zone we first made analytical
continuation for the matrix self-energy from Matsubara frequencies to the real
axis using the Pade approximation \cite{pade}, and then numerically inverted 
 the Green-function matrix as in Eq. (\ref{Gk})
for each $\mathbf{k}$-point.
In the self-consistent solution of the FLEX equations we used 1024 Matsubara
frequencies and the FFT-scheme with the energy cut-off at 100 eV. The sum over
irreducible Brillouin zone have been made with 72 k-points for SCF-iterations
and with 1661 k-points for the final total density of states.

The depolarization of states near the Fermi level is
another important correlation effect. 
The decrease of the ratio $P=\left[ N_{\uparrow }\left(
E_F\right) -N_{\downarrow }\left( E_F\right) \right] /\left[ N_{\uparrow
}\left( E_F\right) +N_{\downarrow }\left( E_F\right) \right] $  is a
typical sign of spin-polaron effects \cite{IKT,uspekhi}.
In our approach this effects  are
taken into account through the $W_{\uparrow \downarrow }^{(ph)}$
terms in the effective spin-polarized LDA++ potential (Eq. (\ref{wpp})).

\begin{figure}
\includegraphics[width=.5\textwidth]{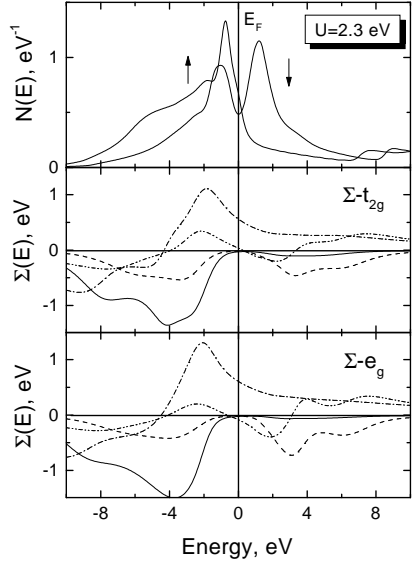}
\vskip 0.5 cm
\caption{
Total spin-polarized density of states and d-part of self-energy for iron 
with  U=2.3 eV and J=0.9 eV for the
temperature T=750 K. Two different self-energies for t$_{2g}$ and
e$_g$ d-states in the cubic crystal field symmetry are presented 
and  four different lines corresponds to imaginary part spin-up (full line)
and spin-down (dashed line) as well as real part spin-up (dashed-dot line)
and spin-down (dashed-double-dot line).
}
\label{SIG}
\end{figure}

The energy dependence of self-energy in Fig.\ref{SIG} shows characteristic features
of moderately correlated systems. At low energies $\left| E\right| <1$ eV we
see a typical Fermi-liquid behavior $Im\Sigma \left( E\right) \sim -E^2,$ $%
\partial Re\Sigma \left( E\right) /\partial E<0.$ At the same time, for the
states beyond this interval within the $d$-bands the damping is rather large
(of the order of 1 eV) so these states corresponds to ill-defined
quasiparticles, especially for occupied states. This is probably one of the
most important conclusions of our calculations. Qualitatively it was already
pointed out in Ref. \cite{treglia} on the basis of a model second-order
perturbation theory calculations. We have shown that this is the case
of realistic quasiparticle structure of iron with the reasonable value of Coulomb
interaction parameter.

Due to noticeable broadening of quasiparticle states the description of the
computational results in terms of effective band structure (determined, for
example, from the maximum of spectral density) would be incomplete. We
present on the Fig.\ref{Ak2} the \textit{full} spectral density $A_\sigma \left( 
{\mathbf k}, E\right) $ including both coherent and incoherent parts as a
function of $\mathbf{k}$ and $E$. We see that in general the maxima of the
spectral density (dark regions)
coincide with the experimentally obtained band structure.
However, for occupied majority spin states at about -3 eV the distribution
of the spectral density is rather broad and the  description 
of this states in terms
of the quasiparticle dispersion is problematic. This conclusion is in
complete quantitative agreement with raw experimental data on angle-resolved
spin-polarized photoemission \cite{kisker} with the broad non-dispersive
second peak in the spin-up spectral function around -3 eV.

\begin{figure}
\begin{minipage}{7.5cm}
\vskip  0cm
\includegraphics[width=.7\textwidth]{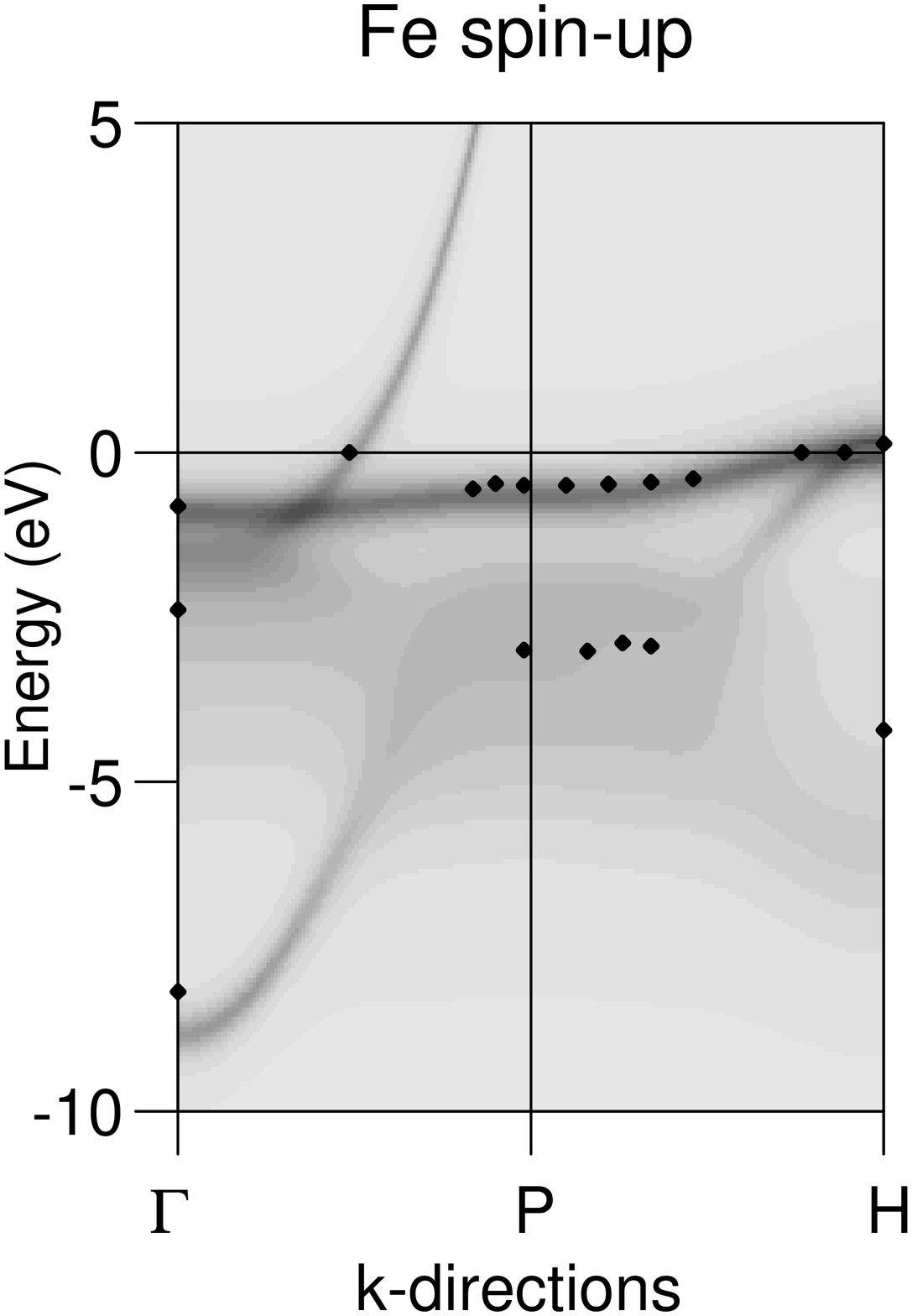}
\end{minipage}
\hspace{-1.0cm}
\begin{minipage}{7.5cm}
\includegraphics[width=.7\textwidth]{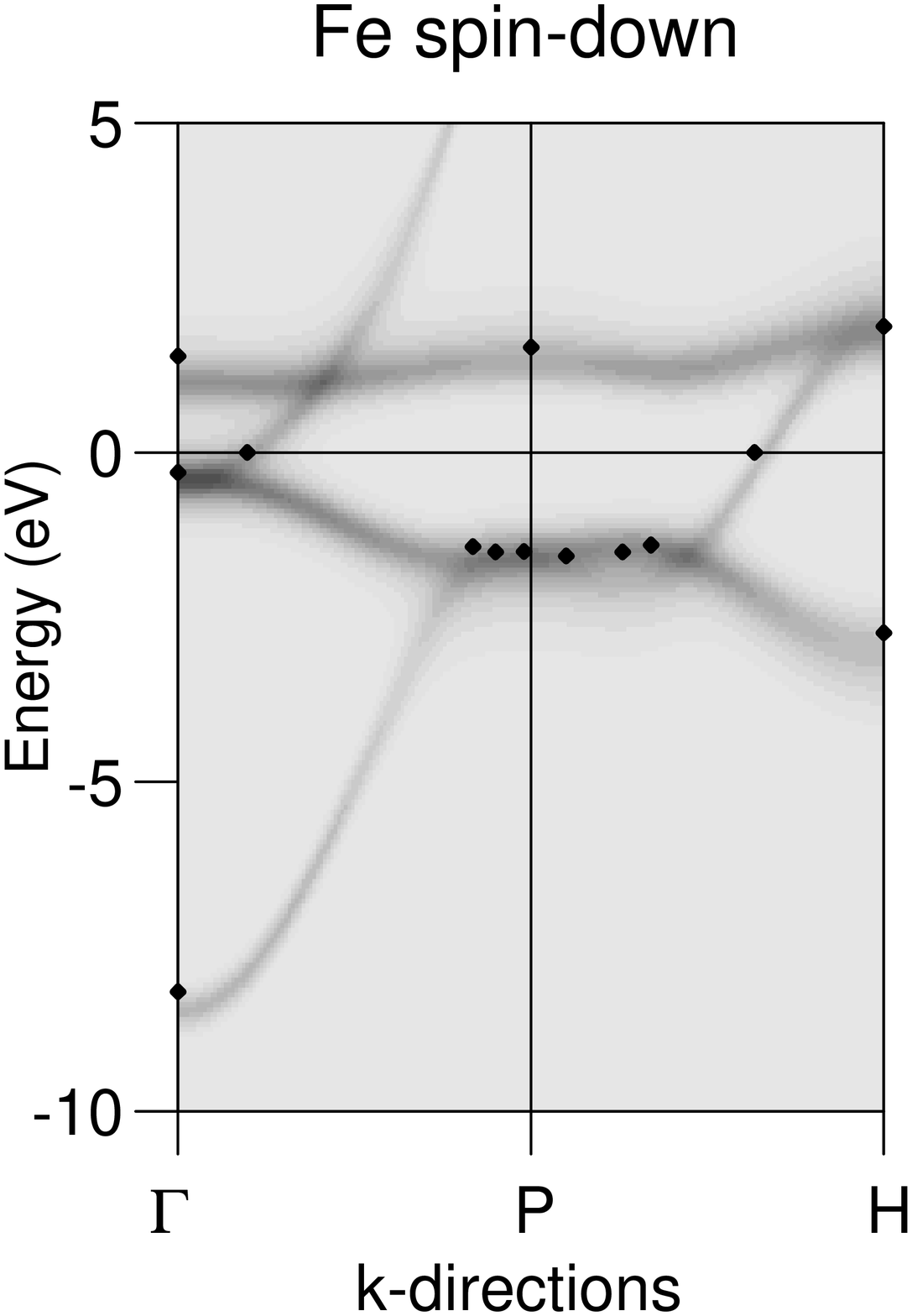}
\end{minipage} \\
\vskip 0.0cm
\caption{
Spectral function of ferromagnetic iron for spin-up (a) and spin-down (b)
and the two k-directions in the Brillouin zone compare with the experimental
angle resolved photoemission and de Haas - van Alphen (at the E$_F$=0)
points 
}
\label{Ak2}
\end{figure}

Comparison of the DOS for transition metals in the Fig. \ref{3dDOS} 
shows an interesting correlations effects. First of all, the most 
prominent difference from the LSDA calculation is observed for the
antiferromagnetic fcc-Mn. There is clear formation of the
lower and upper Hubbard bands around $\pm$ 3 eV. 
Such behavior is related with the
half-field Mn d-shell which corresponds to a large phase space for particle-hole fluctuations.
For the ferromagnetic bcc-Fe the p-h excitations are suppressed by the large
exchange splitting and a bcc structural minimum in the DOS near the Fermi level. 
In the case of ferromagnetic fcc-Co and Ni the correlation effects are more
important then for Fe, since there is no structural bcc-dip in the density
of states. One could see a formation of a "three-peak" structure for the
spin-down DOS for Co and Ni and satellite formation around -5 eV.
In order to describe the satellite formation more carefully one need to
include T-matrix effects \cite{liebsch,drhal}  or use the QMC scheme in LDA+DMFT
calculations.
Finally, there is no big correlation effects in non-magnetic fcc-Cu,
since the d-states are located well bellow the Fermi level.

\begin{figure}
\includegraphics[width=.6\textwidth]{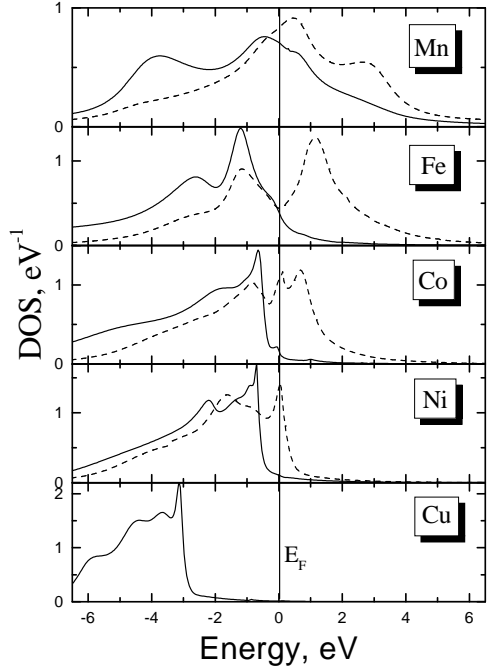}
\caption{
The spin-polarized partial 3d density of states for  
different transition metals at the
temperature T=750 K. 
The full line is the spin-up DOS 
and the dashed line is the spin-down DOS. 
}
\label{3dDOS}
\end{figure}

Using the self-consistent values for $\Sigma _m(i\omega )$ computed
by QMC technique (Section II) we calculate the
exchange interactions (Eq.\ref{Jij}) and spin-wave spectrum (Eq.\ref{om})
using the four-dimensional fast Fourier transform (FFT) method \cite{FFT} for
$({\bf k},i\omega )$ space with the mesh $20^3\times 320$. 
The spin-wave spectrum for ferromagnetic iron is presented in Fig.\ref{SW}
in comparison with the results of LSDA-exchange calculations \cite{LKG} and with
different experimental data \cite{omexp1,omexp2,omexp3}.
This room-temperature neutron scattering experiments has a sample
dependence (Fe-12\%Si in Ref. \cite{omexp1,omexp3} and Fe-4\%Si
in Ref. \cite{omexp2}) due to problems with the bcc-Fe crystal growth. 
Note that for
high-energy spin-waves the experimental data \cite{omexp3} has large
error-bars due to Stoner damping (we show one experimental point with the
uncertainties in the $\bf q$ space). On the other hand, the expression of
magnon frequency in terms of exchange parameters itself becomes problematic
in that region due to breakdown of adiabatic approximation, as it is
discussed above.
 Therefore we think that comparison of theoretical results with experimental
spin-wave spectrum for the large energy needs additional investigation of Stoner
excitation and required calculations of dynamical
susceptibility in the LDA++ approach \cite{dinf}. 
Within the LSDA
scheme one could  use the linear-response formalism \cite{Savr} to calculate
the spin-wave spectrum with the Stoner renormalizations, which should gives
in principle the same spin-wave stiffness as our LSDA calculations. 
Our LSDA spin-wave spectrum agree well with the results of
frozen magnon calculations \cite{Sandratskii,Halilov}.

\vskip 1.0cm 
\begin{figure}[tbp]
\includegraphics[width=.4\textwidth]{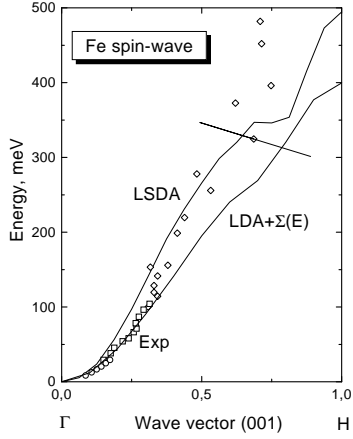}
\vskip 0.5cm
\caption{The spin-wave spectrum for ferromagnetic iron in the LSDA and LDA+$%
\Sigma$ approximations compared with different experiments (circles [16],
squares [17], and diamonds [18]) (a); The corresponding spin-wave spectrum
from LDA+$\Sigma$ scheme in the (110) plane (b).}
\label{SW}
\end{figure}

At the lower-energy,
where the present adiabatic theory is reliable,
the LDA++ spin-waves spectrum  agree better with the
experiments then the result of the LSDA calculations. 
Experimental
value of the spin-wave stiffness D=280 meV/A$^2$ \cite{omexp2} agrees well with
the theoretical LDA++ estimations of 260 meV/A$^2$ \cite{exchlda}.

\appendix

\section*{Appendix}

Here we prove that the expression for the stiffness constant 
Eq.(\ref{D21}) is exact in the
framework of DMFT scheme. 

A rigorous expression for the stiffness constant has been obtained in
Ref. \cite{HE}:

\begin{eqnarray}
2<S^{z}>D_{\alpha \beta }q_{\alpha }q_{\beta }=C({\bf q})-i{\bf q} \cdot \int
\frac{%
d^{4}k}{(2\pi )^{4}}{\pmb \gamma }(k\downarrow ,k\uparrow )G_{\downarrow
}(k)G_{\uparrow }(k)({\bf v}_{{\bf k}}\cdot {\bf q)}
\label{ap1}
\end{eqnarray}
where

\[
C({\bf q})=\frac{1}{2}q_{\alpha }q_{\beta }Tr_{L}\sum_{{\bf k}}\frac{%
\partial ^{2}t({\bf k})}{\partial k_{\alpha }\partial k_{\beta
}}(n_{{\bf k}%
\uparrow }+n_{{\bf k}\downarrow }),
\]
$k=({\bf k}, k_0)$ is a momentum-energy 4-vector, $G(k)$ is the electron
Green function, $t({\bf k})$ is the one-particle (band) Hamiltonian,
$n({\bf k})$ is the one-particle distribution function, ${\pmb \gamma }$
is the
irreducible vector three-leg vertex.
The irreducible scalar and vector vertices, $\gamma$ and ${\pmb \gamma}$,
are connected with the
total ones, $\Gamma_0$ and ${\bf \Gamma }$, by Bethe-Salpeter equations
and the total
vertices satisfy the Ward-Takahashi identity
which is the consequence of the rotational invariance.
Taking into account the Bethe-Salpeter equations one can rewrite them
in terms of the irreducible vertices (for more details, see
\cite{HE}).
Let us assume  now that the scalar irreducible vertex
$\gamma $ and $\Sigma $ are momentum-independent
(which is
the real case in DMFT approach).
Considering the ${\bf q}$-dependent part of the Ward-Takahashi identity
in the limit ${\bf q\longrightarrow 0}$
we
have:
$
{\pmb \gamma }(k \downarrow ,k\uparrow )=\frac{\partial t({\bf k})}{%
\partial {\bf k}}\equiv {\bf v}({\bf k}).
$
Substituting this into Eq.(\ref{ap1})
one has

\begin{eqnarray}
D_{\alpha \beta }=\frac{1}{2<S^{z}>}Tr_{L}\{\sum_{{\bf k}}\frac{\partial
^{2}t({\bf k})}{\partial k_{\alpha }\partial k_{\beta }}n_{{\bf
k}}-i\int
\frac{d^{4}k}{(2\pi )^{4}}v_{{\bf k}\alpha }G_{\downarrow }(k)v_{{\bf
k}%
\beta }G_{\uparrow }(k)\}
\label{ap2}
\end{eqnarray}

On the other hand, our Eq.(\ref{om}) reads
$
2<S^{z}>D_{\alpha \beta }=\frac{1}{2}\frac{\partial ^{2}J({\bf
q})}{\partial
q_{\alpha }\partial q_{\beta }}|_{{\bf q}\longrightarrow 0}
$
where $J({\bf q})$ is defined by Eq.(\ref{Jij}). Calculating the
derivatives of the
exchange parameters we obtain 

\begin{eqnarray*}
\frac{\partial J({\bf q})}{\partial q_{\alpha }} &=&i\int
\frac{d^{4}k}{(2\pi )^{4}}Tr_{L}S(k)G_{\downarrow
}(k)\frac{\partial
t({\bf k})}{\partial k_{\alpha }}G_{\downarrow }(k)S(k)G_{\uparrow
}(k+q)
\end{eqnarray*}

(we shift here $k\longrightarrow k+q$ in the integrand). Then

\[
\frac{\partial ^{2}J({\bf q})}{\partial q_{\alpha }\partial q_{\beta
}}|_{%
{\bf q}\longrightarrow 0}=i\int \frac{d^{4}k}{(2\pi )^{4}}%
Tr_{L}S(k)G_{\downarrow }(k)\frac{\partial t({\bf k})}{\partial
k_{\alpha }}%
G_{\downarrow }(k)S(k)G_{\uparrow }(k)\frac{\partial t({\bf
k})}{\partial
k_{\beta }}G_{\uparrow }(k)
\]

This expression can be simplified by using the sum rule
$
(G_{\uparrow }-G_{\downarrow })=G_{\uparrow }SG_{\downarrow }
$
(where $S=\Sigma _{\uparrow }-\Sigma_{\downarrow } $) which is the
consequence
of the Dyson equation provided that only $\Sigma$ is spin-dependent.
Taking it into account in Eq.(\ref{ap2}) one has

\[
\frac{\partial ^{2}J({\bf q})}{\partial q_{\alpha }\partial q_{\beta }}|_{%
{\bf q}\longrightarrow 0}=i\int \frac{d^{4}k}{(2\pi )^{4}}%
Tr_{L}\{2G_{\uparrow }(k{\bf )}\frac{\partial t({\bf k})}{\partial k_{\alpha
}}G_{\downarrow }(k)\frac{\partial t({\bf k})}{\partial k_{\beta }}%
+\sum_{\sigma }G_{\sigma }^{2}(k{\bf )}\frac{\partial t({\bf k})}{\partial
k_{\alpha }}\frac{\partial t({\bf k})}{\partial k_{\beta }}\}
\]

The first term is exactly coincide with the last one in Eq.(\ref{ap1}) and the
first term
can be transformed further using the identity:
$
G_{\sigma }^{2}(k)=\frac{\partial G_{\sigma }(k{\bf )}}{\partial t({\bf
k})}
$
Then
\[
\int \frac{d^{4}k}{(2\pi )^{4}i}\sum_{\sigma }G_{\sigma }^{2}(k{\bf
)}\frac{%
\partial t({\bf k})}{\partial k_{\alpha }}\frac{\partial t({\bf k})}{%
\partial k_{\beta }}=\sum_{{\bf k}\sigma }\frac{\partial n_{\sigma
}({\bf k)}%
}{\partial t({\bf k})}\frac{\partial t({\bf k})}{\partial k_{\alpha
}}\frac{%
\partial t({\bf k})}{\partial k_{\beta }}
\]
Since $n_{\sigma }({\bf k)=}n_{\sigma }[t({\bf k)]}$ we have:
$
\frac{\partial }{\partial k_{\alpha }}n_{\sigma }[t({\bf
k)]=}\frac{\partial
n_{\sigma }({\bf k)}}{\partial t({\bf k})}\frac{\partial t({\bf k})}{%
\partial k_{\alpha }}
$
and finally, integrating by part, one obtains:

\[
-\sum_{{\bf k}}\frac{\partial n_{\sigma }}{\partial k_{\alpha }}\frac{%
\partial t({\bf k})}{\partial k_{\beta }}=\sum_{{\bf k}}n_{\sigma }({\bf
k)}%
\frac{\partial ^{2}t({\bf k})}{\partial k_{\alpha }\partial k_{\beta }}
\]
Thus, our expression (Eq.(\ref{D21})) coincides with the exact one (Eq.(\ref{ap1})). We use
here the
only assumption that both the self-energy and three-leg irreducible
vertex are
momentum independent as well as the Ward-Takahshi identities which are
exact consequences of the rotationally invariance of the spin system.

\section{Conclusions}

We have proposed a general scheme for investigation of the correlation effects
in the quasiparticle band structure calculations for itinerant-electron
magnets. This approach is based on the combination of the dynamical mean-field theory and
the fluctuating exchange approximation. 
Application of LDA+DMFT method  gives an adequate
description of the quasiparticle electronic structure for ferromagnetic iron.
The main
correlation effects in the electron energy spectrum are strong damping of
the occupied states below 1 eV from the Fermi level $E_F$ and essential
depolarization of the states in the vicinity of $E_F$. We obtained a
reasonable agreement with different experimental spectral data 
(spin-polarized photo- and thermoemission). The method is rather
universal and can be applied for other magnetic systems, both ferro- and
antiferromagnets.

We discussed as well a general method for the investigation of magnetic
interactions in the correlated electron systems. This scheme is not based on
the perturbation theory in ``$U$'' and could be applied for rare-earth
systems where both the effect of the band structure and the multiplet
effects are crucial for a rich magnetic phase diagram. Our general
expressions are valid in relativistic case and can be used for the
calculation of both exchange and Dzialoshinskii- Moriya interactions, and
magnetic anisotropy \cite{exchlda}. An illustrative example of ferromagnetic iron shows
that the correlation effects in exchange interactions may be noticeable even
in such moderately correlated systems. For rare-earth metals and their
compounds, colossal magnetoresistance materials or high-$T_c$ systems, this
effect may be much more important. For example, the careful investigations
of exchange interactions in MnO within the LSDA, LDA+U and optimized
potential methods for MnO \cite{Sol} show the disagreement with
experimental spin-wave spectrum (even for small ${\bf q}$), and indicate a
possible role of correlation effects.

This work demonstrates an essential difference
between spin density functional approach and LDA++ formalism. The latter method deals
with the thermodynamic potential as a functional of the local Green function rather
than the electron density. Nevertheless, there is a close connection 
between two techniques (the self-energy corresponds to the exchange- correlation
potential, etc). In particular, an analog of local force theorem can be
proved for LDA++ approach. It may be useful not only for the calculation of
magnetic interactions but also for elastic stresses, in particular,
pressure, and other physical properties.

\section{Acknowledgments}

We are grateful to O.K. Andersen, C. Carbone, P. Fulde,
O. Gunnarsson, G. Kotliar, and A. Georges for helpful discussions.
The work was supported by the Netherlands Organization
for Scientific Research (NWO project 047-008-16)
and partially supported by Russian Basic Research Foundation,
grant 00-15-96544.

%

\end{document}